\documentclass[twocolumn,showpacs,preprintnumbers,amsmath,amssymb]{revtex4}

\usepackage{graphicx}
\usepackage{dcolumn}
\usepackage{bm}

\begin{document}

\preprint{RBRC-748, YITP-08-63}

\title{
Dynamical study of bare $\sigma$ pole
with $1/N_c$ classifications}

\author{Toru Kojo$^{1}$ and Daisuke Jido$^2$}

\affiliation{
$^1$RIKEN BNL Research Center, Brookhaven National Laboratory,
Upton, New York 11973, USA \\
$^2$Yukawa Institute for Theoretical Physics, Kyoto University,
          Kyoto 606-8502, Japan
}

\date{\today}

\begin{abstract}
Systematic $1/N_c$ counting of correlators is performed
to directly relate quark-gluon dynamics
to qualitatively different hadronic states order by order.
Both 2q and 4q correlators of $\sigma$ quanta are analyzed
with $1/N_c$ separation of the instanton, glueball, and in particular, 
two meson scattering states.
The {\it bare} resonance pole 
with no mixing effects are
studied with the QCD sum rules (QSR).
The bare mass relation for large $N_c$ mesons,
$m_{\rho}<m_{4q}^{I=J=0}<m_{2q}^{I=J=0}$, is derived.
The firm theoretical ground of the QSR at the leading $1/N_{c}$
analyses is also emphasized.
\end{abstract}

\pacs{12.39.Mk,11.55.Hx,11.30.Rd}

\maketitle


Quantum chromodynamical (QCD) 
descriptions of hadron properties
have direct relevance to understanding of
the nonperturbative aspects
of the strong interaction. 
For instance,
success of the constituent quark
model for global hadron spectra
has illuminated some properties of 
chiral symmetry breaking and confinement,
and provided the concept of constituent quarks
as quasiparticles inside hadrons.
Yet, not all hadrons are alike. 
In addition to the well-established Nambu-Goldstone bosons,
there exist some exceptional states,
for example,  the light scalar mesons 
($\sigma$, $\kappa$, $a_0$, $f_0$)~\cite{Close}, newly 
observed charmonia ($X$, $Y$, $Z$) \cite{Belle},
 baryon resonance
$\Lambda(1405)$ and 
flavor exotic $\Theta^{+}(1540)$ \cite{LEPS}. 
The studies of these exotic hadrons 
could provide new viewpoints beyond the simple consitituent 
quark picture, such as
multi-quark components and/or inter-hadron-dynamics.

The lightest scalar meson, $\sigma$ with 
$I=J=0$, is a typical example
which has almost all important ingredients
as the exotic hadrons.
The $\sigma$ meson includes not only usual $q \bar q$ (2q)
but hadronic states
beyond the constituent quark picture:
glueball, $\pi \pi$ molecule, and $qq\bar q \bar q$ (4q) state
with diquark correlation \cite{Jaffe}.
Thus it is a good laboratory to explore the properties 
and interplay of these states.
Extraction of these properties
has direct phenomenological importance to
not only the hadron spectroscopy
but also nuclear/quark-hadron mater,
through the properties of nuclear force
~\cite{Johnston}, 
chiral order parameter of QCD~\cite{Hatsuda}, 
and diquark correlation \cite{Jaffe}.

The studies of the $\sigma$ meson, however,
are not straightforward.
Since $\sigma$ can be described as admixture of several 
hadronic states,
it is difficult to identify which hadronic
states are responsible to which part of the $\sigma$ properties.
Experimental information is still not satisfactory
to derive the definite conclusion.
Therefore, as a first step,
it is important to theoretically clarify the properties
of each hadronic state in the absence of mixing
with other states \cite{Lattice} and,
at the same time, to illuminate the role
of interplay between these hadronic states,
by examining the descrepancies between
states with no mixing effects
and the experimental $\sigma$
data with all mixing effects.

For this purpose,
we introduce the classification based on 
inverse expansion of number of color, 
$1/N_{c}$ \cite{tHooft, Witten}, 
for the hadronic states 
in the correlators made of quark-gluon fields:
$\Pi(q^2) = i \int d^{4}x \, 
e^{iq \cdot x}\langle T J(x) J^{\dagger}(0) \rangle$,
%
including contributions from all possible 
hadronic intermediate states with 
the same quantum number as $J(x)$.
One of the largest virtues of $1/N_{c}$ expansion
is that it directly relates the $1/N_c$ classifications 
of the quark-gluon graphs
to the qualitative classifications of the hadronic states
with the same quantum number,
in a way that the mixing of these hadronic states
are suppressed by higher order quark-gluon graphs of $1/N_{c}$.
Then we can concentrate on the graphs 
for the hadronic states of our interest, separating
mixing effects from higher order of $1/N_c$.

In this letter, 
we demonstrate this idea in the case of the 
correlators of 2q and 4q operators 
with the $\sigma$ quantum number.
On the basis of the $1/N_{c}$ distinction,
we can give an inductive definition of 
the bare "2q" and "4q" states, being free from
the contributions of
glueball, instanton,
and, in particular, $\pi\pi$ scattering states,
which are the origin of the large width $\sim 500$ MeV \cite{Caprini}
and large background in the $\sigma$ meson spectrum.
A novel consequence of our approach is
that the bare "4q" state can be investigated
independently of 2-meson states,
and this explicitly demonstrates
the efficiency of the $1/N_{c}$ distinction of 
states with the same quantum number but with
qualitative differences.   
The existence and properties of the "4q" state are dynamically 
studied with comparing them to the "2q" state 
through the QCD sum rules (QSR) \cite{Shifman}, 
whose theoretical ground is firm
in the leading $1/N_{c}$.
We will figure out the importance of the 
"4q" component with smaller mass than "2q" case by
$150 \sim 200$ MeV despite of larger number of quarks
participating in the dynamics.
This indicates the existence of the nontrivial correlation
for the mass reduction of the "4q" system.

The interpolating fields used in this work
are summarized as follows:
The 2q interpolating fields are described as
$J_{M}^{F}=\bar{q} \tau_F \Gamma_M q$,
where Dirac matrix $\Gamma_M$ labeled by $M=(S,P,V,A,T)$ 
for
($1, i\gamma_5, \gamma_{\mu}, \gamma_{\mu}\gamma_5, \sigma_{\mu\nu}$),
respectively, and $\tau_F\ (F=1,2,3)$ are the Pauli matrices 
acting on $q=(u,d)^T$. 
The 4q operators with the $\sigma$ quantum number
are given (assuming the ideal mixing for the $\sigma$ meson)
by $J_{MM}(x) = \sum_{F=1}^3 J_{M}^{F}(x) J_{M}^{F}(x)$
as products of meson operators
(Hereafter we take the SU(2) chiral limit
for simplicity).

Here we first see the $1/N_{c}$ linking 
between quark-gluon dynamics and hadronic states
in the case of the 2q correlators.
The known facts are as follows:
\cite{tHooft,Witten}: 
1) For quark-gluon diagrams, $n$-internal quark loops 
are suppressed by $1/N_{c}^n$.
In terms of hadron,
$n$-meson or multiquark with $(q\bar{q})^n$ production from 
$J_M$ are suppressed by $1/N_{c}^n$.
2) The disconnected diagrams with two gluon emission
are suppressed by $1/N_{c}$. 
In terms of hadron, $q\bar{q}$-glueball mixing is suppressed
by $1/N_{c}$.
3) Instanton effects are suppressed by $\sim e^{-N_{c}}$.
4) For the meson properties 
the $n$-meson couplings are given as $g_{nM}=O(N_{c}^{(2-n)/2})$. 
Here 1)$\sim$3) implies that
leading diagrams with $O(N_{c})$ 
can be naturally interpreted as the bare
"2q" state
since the 4q propagation 
diagrams/$\pi\pi$ scattering, glueball, and instanton
contributions
do not appear at this order.

Similar identification and subsequent separation
are also possible in the case of the 4q correlator,
which incorporates the 4q participating diagrams
from the beginning.
Here we give inductive definition on the "4q" state
following the $1/N_c$ based orthogonal conditions:
a) "4q" can {\it not}
appear in the leading $N_{c}$ 2q correlator,
b) "4q" can appear in the 4q correlators
even after the separation of 2-meson scattering states.
These conditions insist that its dynamical origin is different
from "2q" and 2-meson molecule states
(The glueball and instanton are easily verified
to be higher order in $1/N_{c}$ than those
considered below,
thus we will not discuss them in the following).

\begin{figure}[b]
\vspace{-0.5cm} 
\includegraphics[width=6.0cm, height=3.2cm]{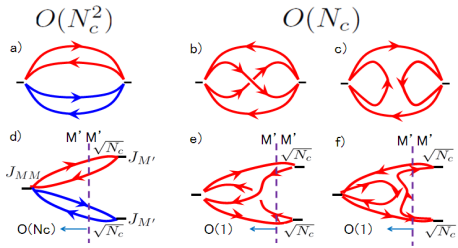} 
\vspace{-0.25cm}
\caption{(Color online) Examples of the
$O(N_c^2)$ and $O(N_c)$ quark-gluon diagrams  
for 2 and 3 point correlators.}
\vspace{-0.3cm}
\label{fig:2pointgraph}
\end{figure}

Although this definition gives a 
convenient starting point to discuss the qualitative 
difference between hadronic states in the
$\sigma$ meson,
the studies of the "4q" component
require systematic $1/N_{c}$ arguments for the 4q correlators,
beyond
the leading $O(N_{c}^2)$ 
quark-gluon diagrams including only 2 planar loops 
(Fig.\ref{fig:2pointgraph},a),
which are naturally interpretted as free 2-meson scattering
and are irrelevant for the studies of the "4q" properties.
Thus we must proceed to the next leading order of $1/N_{c}$,
$O(N_c)$ diagrams
which could include 2-meson scattering, "2q", and "4q" states
at this next leading order of $1/N_{c}$. 

The $O(N_{c})$ quark-gluon diagrams
in the 2-point function
could include the various hadronic contributions.
The easiest way to classify them is to consider
the overlap strength
of the operator $J_{MM}$ with hadronic states,
involving all the elements
of hadronic diagrams.
We first classify the overlap strength of the 4q field 
with the 2 meson states based on $1/N_{c}$,
employing 3-point correlator
among the 4q current $J_{MM}$ 
and two separated meson operators $J_{M'}$
(Fig.\ref{fig:2pointgraph}, d-f).
An explicit examination of quark-gluon graphs 
shows that
the leading order diagrams are $O(N_c^2)$ for $M=M'$ case,
and $O(N_c)$ for $M \neq M'$.
Combining these facts with
the overlap strength of $J_M'$ with the 2q meson state $|M'\rangle$ 
is $O(N_{c}^{1/2}$), the remaining part
should be 
\begin{eqnarray}
\langle 0|J_{MM}|M'M'\rangle = O(N_c)\delta_{MM'} + O(1)+....
\label{eq:overlap}
\end{eqnarray}
The first term represents the direct coupling to
$|MM \rangle$,
while the second term reflects
that transition into $|M'M'\rangle$ final state
needs the additional interactions.
This higher order counting is crucial
for the separation of the $\pi\pi$ scattering
states from 2-point correlators.

On the other hand, 
the overlap strength with "2q" and "4q" states
cannot be deduced from $1/N_c$ arguments only,
and are assumed to be
\begin{eqnarray}
\langle 0|J_{MM}|R\rangle = O(N_c^{1/2}), 
\ \ (R="2q"\ {\rm or}\ "4q")
\label{eq:assumption}
\end{eqnarray}
which will be assured later through the dynamical calculations.
Similarily, the "4q" state will be identified
by examining the quantitative difference of poles in 2q and 4q correlators,
which is found to be large enough 
to distinguish the "4q" and "2q" states.
The coupling of $R$ 
to two mesons is estimated by 3-point function in the same way 
to obtain the meson couplings performed in Refs.~\cite{Witten} and 
is found to be $O(N_{c}^{-1/2})$.

Now we can classify
the hadronic states in
the 2-point correlators $\langle J_{MM} J_{M'M'} \rangle$ 
based on $1/N_{c}$ (See, Fig.\ref{fig:2pointgraph}, a-c):
(i) If $M=M'$, $O(N_c^2)$ quark-gluon graphs include only the free
2M scattering states in the region $E \ge 2m_M$.
Otherwise, the contributions from these quark-gluon diagrams vanish,
indicating absence of 2 meson scattering states.
(ii) $O(N_c)$ graphs include the 2M or 2M' scattering
and possible resonance, "2q" and/or "4q".

Note that the relations (\ref{eq:overlap}) in the case of
$M,M'\neq P\ {\rm nor}\ A$ indicate that
the 2$\pi$ scattering states are not included up to $O(N_c)$ diagrams,
and then the resonance peaks (if exist) {\it below} $2m_{M}$
are isolated and have zero width since
the decay channel is absent.
Therefore,
now we can reduce the $\sigma$ spectrum in the 4q correlator
into peak(s) plus continuum  
{\it if we retain only diagrams up to $O(N_c)$}.

This separate investigation of the $O(N_c^2)$ and $O(N_c)$ part
of QCD dynamics enables to perform
the step by step analyses for
the 2 meson scattering, "2q", and "4q" spectra.
In the application of the QSR,
we perform the operator product expansion (OPE)
for the correlators in deep Euclidean region ($q^2=-Q^2$),
then translate them, {\it term by term of $1/N_c$},
into the {\it integral} of the hadronic 
spectral function through the
dispersion relation:
\begin{eqnarray}
{\rm \Pi}^{ope}_{N_c^n}(-Q^2) 
= \int_0^{\infty}\hspace{-0.2cm}
ds\ \frac{1}{\pi}\frac{ {\rm Im \Pi}^h_{N_c^n}(s) }{s+Q^2}
\ \ (n=2,1).
\label{disp}
\end{eqnarray}

Now we emphasize the practical aspects of
$1/N_c$ expansion in the application of the QSR.
First, the higher dimension condensates in the OPE,
whose values have been not well-known
despite of their importance, can be factorized into
the products of known condensates,
$\langle \bar q q \rangle$, $\langle G^2 \rangle$, and
$\langle \bar q g_s \sigma G q \rangle$.
For example,
$\langle (\bar q Q) 
(\bar Q q) \rangle \nonumber
= \langle \bar q Q \rangle 
\langle \bar Q q \rangle 
+ \sum \langle 0| \bar q Q |M \rangle 
\langle M| \bar Q q |0 \rangle
\rightarrow \langle \bar q Q \rangle 
\langle \bar Q q \rangle$,
holds in leading $1/N_c$ estimation
since $\langle \bar q q \rangle$ is $O(N_c)$,
while $\langle 0|\bar q \Gamma q |M \rangle$ is $O(N_c^{1/2})$
\cite{Witten}.
Keeping this merit,
we will deduce the final $O(N_c)$ results from
the off diagonal correlator $\langle J_{VV}J^{\dag}_{SS} \rangle$,
whose leading order is $O(N_c)$
thus without the factorization violations
at the $O(N_c)$ OPE. 

Secondly, the lowest resonance in the reduced
$O(N_c)$ spectra ${\rm Im \Pi}_{N_c}^{h}(s)$
for the "2q" and "4q" states,
can be described as the sharp peak 
because of the absence of the decay channel.
Applying the usual quark-hadron duality approximations
to the higher excited states,
$\pi{\rm Im \Pi}^{h}_{N_c} (s) = \lambda^2 \delta(s-m_h^2)
+ \theta(s-s_{th}) \pi {\rm Im \Pi}_{N_c}^{ope} (s)$,
and after the Borel transformation for Eq.(\ref{disp}),
we can express the effective mass as
\begin{eqnarray} 
m_h^2(M^2;s_{th}) \equiv \frac{ \int_0^{s_{th}} \!ds
\ e^{-s/M^2 }s\ {\rm Im} \Pi^{ope} (s) }
 { \int_0^{s_{th}} \!ds\ e^{-s/M^2 }{\rm Im} \Pi^{ope} (s) }.
 \label{eq:peakmass}
\end{eqnarray}
$s_{th}$ can be uniquely fixed to satisfy the
least sensitivity \cite{LScriteria} of the expression 
(\ref{eq:peakmass})
against the variation of $M$,
since the physical peak should not depend on the
artificial expansion parameter $M$.
This criterion is justified only when the peak
is very narrow,
and our $1/N_c$ reduction of spectra 
is essential for its application
to allow the QSR framework to determine all physical parameters
($m_h, \lambda, s_{th}$)
in self-contained way.

In practical application of the QSR,
it is essential 
to reduce errors of finite order trunction of OPE
and of the quark-hadron duality approximation.
Thus Eq.(\ref{eq:peakmass}) must be evaluated in the appropriate
$M^2$ window ($M^2_{min},\ M^2_{max}(s_{th})$)
to achieve the conditions: good OPE convergence for $M_{min}$
(highest dimension terms $\le$ 10\% of whole OPE)
and sufficient ground state saturation for $M_{max}$ 
(pole contribution $\ge$ 50\% of the total) \cite{Reinders,KHJ,KJ}.
Without the $M^2$ constraint,
we are often stuck with the {\it pseudo-peak} artifacts
\cite{KJ} often seen in multiquark SRs.
Thus we carry out OPE up to dimension 12
\cite{drop} to include the sufficient low energy contributions
which is essential to find the reasonable $M^2$ window
\cite{KJ,KHJ}.

We summarize the numerical values used in the analyses.
The gauge coupling constant behaves like $O(N_c^{-1/2})$, 
and the condensates, $\langle O \rangle=$($\langle \bar{q}q\rangle$,
$\langle \alpha_s G^2 \rangle$, 
$\langle \bar q g_s \sigma G q \rangle$)
are $O(N_c)$.
Here we additionaly put simple $N_c$ 
scaling assumptions:
$\alpha_s |_{N_c} = 3\alpha_s/N_c$,
$\langle O \rangle |_{N_c} = \langle O \rangle N_c/3$.
We use the following values with errors for the $N_c=3$ case,
$\alpha_s({\rm 1GeV}) =0.4$,
 $\langle \alpha_s G^2/\pi \rangle = (0.33\ {\rm GeV})^4$,
$\langle \bar{q}q\rangle=-(0.25 \pm 0.03\ {\rm GeV})^3$,
and 
$\langle \bar q g_s \sigma G q \rangle/ 
\langle \bar{q}q\rangle
= (0.8 \pm 0.1)\ {\rm GeV^2}$.
The results shown below will be obtained with
the central values.
We finally show $\langle \bar q q \rangle$ and $m_0^2$
dependence of masses in a wide range 
since most of errors come from
these values.

\begin{figure}[t]
\vspace{-0.1cm}
\includegraphics[width=8.0cm]{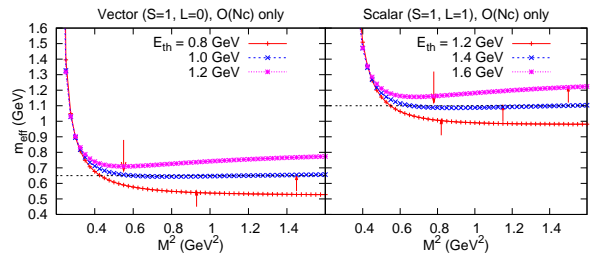}
\vspace{-0.3cm}
\caption{(Color online) 
The $O(N_{c})$ 
effective mass plots for vector and scalar mesons
in the cases of various $E_{th}$ values.
The downward (upward) arrow
represent the $M_{min}^2$ ($M_{max}^2(s_{th})$).}
\label{fig:2qmeson}
\vspace{-0.5cm}
\end{figure}

We start the Borel analyses from the case of
the large $N_c$ 2q correlators
for the vector meson as a reference and the scalar meson as the "2q" state
in the $\sigma$ meson.
The OPE results (up to dimension 6) are nothing but
results in the literature with employing the factorization.
Shown in Fig.\ref{fig:2qmeson} are
the effective mass for the large $N_c$ vector and scalar meson 
as functions of $M^2$ for various $E_{th}$.
The downarrow (upperarrows) indicates the value of 
$M^2_{min}$ ($M^2_{max}(s_{th})$).
Following the $E_{th}$ ($\equiv \sqrt{ s_{th} }$) fixing criterion,
we obtain $E_{th}$ to 1.0 (1.4) GeV for the vector (scalar) meson,
and determine the mass as 0.65 (1.10) GeV.
The mass splitting $\sim$0.45 GeV between the vector and scalar mesons
roughly
coincides with the angular excitation energy expected from the 
naive consitituent quark picture.   
The reasons for slightly small value of the large $N_c$
$\rho$ meson mass
could be due to the absence of the factorization violations
\cite{violation}
if the large $N_c$ scaling of condensates holds.
Since
our original interests are in the qualitative
difference in large $N_c$ mesons
rather than their absolute values,
thus we will not discuss the details of the absolute values further.

Now we turn to the 4q correlator results of our main interest.
We have investigated $O(N_c^2)$ and $O(N_c)$ part of 
$\langle J_{SS}J^{\dag}_{SS} \rangle$,
$\langle J_{VV}J^{\dag}_{VV} \rangle$,
and $O(N_c)$ part of
$\langle J_{VV}J^{\dag}_{SS} \rangle$.
We have checked that the typical effective masses for $O(N_c^2)$
part of $\langle J_{SS} J^{\dag}_{SS} \rangle$
($\langle J_{VV} J^{\dag}_{VV} \rangle$)
are well above the twice of
the large $N_c$ meson mass 2.2 (1.3) GeV,
indicating that there is no prominent structure
below the free 2-meson threshold,
as expected from $1/N_c$ arguments.

\begin{figure}[t]
\includegraphics{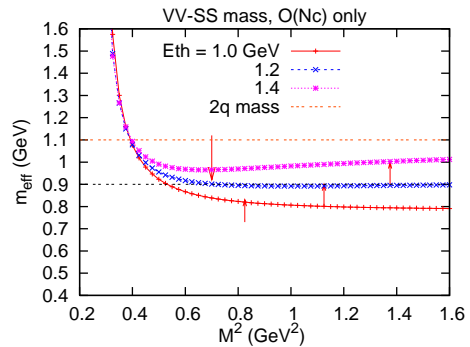}
\vspace{-0.3cm}
\caption{(Color online) 
The effective mass plots for 
$\langle J_{SS} J^{\dag}_{VV} \rangle$,
including only $O(N_c)$ OPE diagrams.
The large $N_c$ "2q" scalar meson mass
is also indicated as a reference.}
\vspace{-0.5cm}
\label{fig:ncspectra}
\end{figure}

The "4q" state can appear from $O(N_c)$ part. 
Shown in Fig.\ref{fig:ncspectra}
are the effective masses deduced from 
$\langle J_{VV} J^{\dag}_{SS} \rangle$
for $E_{th}$=1.0, 1.2, and 1.4 GeV.
We take the $E_{th}$=1.2 GeV case
and evaluate its mass as $\sim$0.90 GeV,
which is obviously lower than that of the "2q" scalar meson case, 
$\sim$1.10 GeV 
in large $N_c$ limit,
and thus is considered as the mass of the "4q" state.
The threshold value 1.2 GeV for $E_{th}$, much below
the 2 scalar (vector) meson threshold, 2.2 (1.3) GeV,
may be due to the "2q" scalar meson contribution,
since our $O(N_c)$ 4q correlators 
can also include the "2q" contribution.
We have also investigated the $O(N_c)$ part of
$\langle J_{SS}J^{\dag}_{SS} \rangle$
($\langle J_{VV}J^{\dag}_{VV} \rangle$),
and obtained the almost same mass 0.80 (0.90) GeV
although they could suffer from the factorization violation
coming from $O(N_c^2)$ OPE.
These 3-independent $O(N_c)$ correlator analyses consistently suggest 
existence of the "4q" state lighter than the "2q" state.

\begin{figure}[t]
\vspace{-0.2cm}
\includegraphics{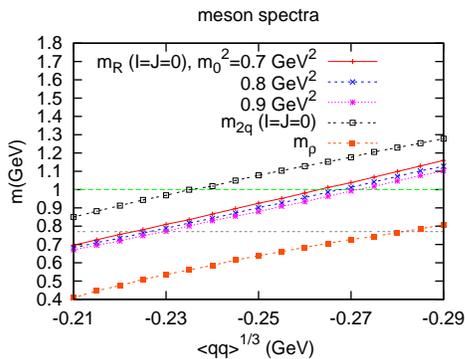}
\vspace{-0.3cm}
\caption{(Color online) 
The condensate value dependence of masses of the large $N_c$ mesons,
"4q" ($m_0^2$=0.7, 0.8, 0.9 GeV$^2$),
"2q" scalar and vector states. 
}
\vspace{-0.6cm}
\label{fig:mesonspectra}
\end{figure}
Finally, we derive a conservative conclusion
which does not depend on the details
of our numerical parameters, especially
$\langle \bar{q}q \rangle$ and $m_0^2$
(The dependence on the other parameters is relatively small). 
Shown in Fig.\ref{fig:mesonspectra} are
"2q" vector, scalar meson masses,
and of the "4q" mass (deduced from 
$\langle J_{VV}J^{\dag}_{SS} \rangle$)
as functions of the $\langle \bar{q}q \rangle$
and $m_0^2$. 
We found that the inequality 
$m_{\rho}<m_{4q}^{I=J=0}<m_{2q}^{I=J=0}$
holds irrespective of details of the condensate values.

The results obtained here
suggest that
the $\sigma$ meson has the 
"4q" component, which is not generated
from $\pi\pi$ interactions but
from those at the quark-gluon level.
Here we comment 
on Pelaez's elaborated work for the $\sigma$ meson
using the unitarized chiral perturbation
with $1/N_c$ expansion \cite{Pelaez}.
He showed that the $\sigma$ state as the pole in the
$\pi\pi$ scattering amplitudes of
T-matrix disappears in large $N_c$ limit 
in contrast to the case of ordinary mesons
such as the $\rho$ meson.
This is not contradicted with our results
since the $\pi\pi$-"4q" mixing 
is suppressed by $1/N_c$
and "4q" state is not accessible 
from the $\pi\pi$ initial states.
This is in sharp contrast to 
the 4q correlator approach
including the "4q" state directly
generated from 4q current.

We conjecture that
the "4q" component may play important role
as a building block of the $\sigma$ meson.
To develop this possiblity,
we plan to study the 3-point correlator 
for the "4q"-$\pi\pi$ coupling strength.
The coupling strength
should be large since the $\sigma$ in nature 
is the broad resonance.
If this is indeed the case, 
the $\sigma$ meson could be described 
as the 4q core clothed by
the $\pi\pi$ clouds. 
The relative importance of the "4q" state
in the $\sigma$ meson can be
investigated through the coupled channel analyses
using the effective Lagrangian including
not only the $\pi$ field but also an
elementary "4q" field,
whose effects are considered to 
be hidden in the parameters or regulation constants
in the usual chiral perturbation approaches.
This is somehow related to the recent arguments
for the $N^*$(1535) resonance \cite{Hyodo2}.

1/$N_c$ arguments developed in this work
are expected to have the wide applications.
First, the spectrum reduction
is applicable to the analyses of
other tetraquark candidates.
We have already obtained results that
the effective mass in $I=2,J=0$ channel 
does not show any stability against $M^2$ variation,
which indicates absence of "4q" states
consistently with experiments.
Second, the QCD sum rules which are 
firm in the large $N_c$ limit
could provide useful information to models based on 
the Gauge/Gravity duality (large $N_c$ QCD $\leftrightarrow$ SUGRA),
through the properties of the large $N_c$ mesons.
All these issues will be reported in future \cite{future}.

We thank Profs. T. Kunihiro and H. Suganuma
for useful discussions and encouragements.
T.K is indebted to
Profs. M. Harada, W. Weise, B. Mueller
for useful discussions during {\it New Fronteers in QCD}
held at YITP,
and also to Profs. T. Hatsuda
and S. Sasaki for several important comments.
We appriciate Prof. D. Kharzeev for
carefully reading the manuscript.
This work is supported by RIKEN, Brookhaven National
Laboratory and the U. S. Department of Energy 
[Contract No. DE-AC02-98CH10886], and by the 
Grant for Scientific Research (No. 20028004)
in Japan.


\begin{thebibliography}{12}
\bibitem{drop} The graphs with multi-gluon condensates
   are usually neglected due to the strong suppression
   by the extra loop factor compared to the
   graphs with quark condensates.
\bibitem{LScriteria} More quantitatively,
 	 within the region $(M_{min}^2, M^2_{max})$,
         we calculate the variance of mass as function of $s_{th}$.
	 Then we select $s_{th}$ which minimizes the variance.
\bibitem{violation} 
	To adjust the $\rho$ meson mass to the experimental value,
	one must include the factor $\sim 2$ factorization violation
	or adopt the value 
	$\langle \bar{q} q \rangle \sim -(0.280\ {\rm GeV})^3$.
\end{thebibliography}
\end{document}